\documentclass{eptcs}
\providecommand{\event}{TFPIE} % Name of the event you are submitting to

\usepackage{tikz}
\usepackage{amsmath}
\usepackage{color}
\usepackage{fancyvrb}

\newcommand{\VERB}{\Verb[commandchars=\\\{\}]}
\DefineVerbatimEnvironment{Highlighting}{Verbatim}{commandchars=\\\{\}}
\newcommand{\KeywordTok}[1]{\textcolor[rgb]{0.00,0.44,0.13}{\textbf{{#1}}}}
\newcommand{\DataTypeTok}[1]{\textcolor[rgb]{0.56,0.13,0.00}{{#1}}}
\newcommand{\DecValTok}[1]{\textcolor[rgb]{0.25,0.63,0.44}{{#1}}}

\newcommand{\SpecialCharTok}[1]{\textcolor[rgb]{0.25,0.44,0.63}{{#1}}}

\newcommand{\FunctionTok}[1]{\textcolor[rgb]{0.02,0.16,0.49}{{#1}}}

\newcommand{\OperatorTok}[1]{\textcolor[rgb]{0.40,0.40,0.40}{{#1}}}

\newcommand{\NormalTok}[1]{{#1}}

\newcommand{\darkgreen}[1]{\textcolor[rgb]{0.00,0.80,0.00}{#1}}

\newcommand{\ket}[1]{\left| #1 \right\rangle}
\newcommand{\bra}[1]{\left\langle #1 \right|}
\newcommand{\abs}[1]{\left| #1 \right|}

\title{Learn Quantum Mechanics with Haskell}
\author{Scott N. Walck
\institute{Department of Physics\\
  Lebanon Valley College\\
  Annville, Pennsylvania, USA}
\email{walck@lvc.edu}
}

\def\titlerunning{Learn Quantum Mechanics with Haskell}
\def\authorrunning{Scott N. Walck}

\begin{document}

\maketitle

\begin{abstract}
To learn quantum mechanics, one must become adept in the use
of various mathematical structures that make up the theory;
one must also become
familiar with some basic laboratory experiments that the theory is designed
to explain.
The laboratory ideas are naturally expressed in one language,
and the theoretical ideas in another.
We present a method for learning quantum mechanics that
begins with a laboratory language for the description and
simulation of simple but essential laboratory experiments,
so that students can gain some intuition about the phenomena that a theory
of quantum mechanics needs to explain.
Then, in parallel with the introduction of the mathematical
framework on which quantum mechanics is based,
we introduce a calculational language for describing
important mathematical objects and operations,
allowing students to do calculations in quantum mechanics, including
calculations that cannot be done by hand.
Finally, we ask students to use the calculational language
to implement a simplified version of the laboratory language,
bringing together the theoretical and laboratory ideas.
\end{abstract}

\section{Introduction}

The theories of twentieth-century physics employ mathematical objects
that are quite removed from our everyday experience of the world
and surprisingly removed from the description of the experiments
that led to or provided evidence for those theories.
Certainly theoretical concepts have motivated and guided
experiments---experimental design is awash in theory---but
if we consider the simplest description of an experiment,
as a chef might write a recipe for a lay cook, the language
would not include references to the abstract
objects that structure the theorist's calculations.

We focus in this paper on the theory of quantum mechanics,
and in particular on the behavior of spin-1/2 particles,
some of the very simplest quantum systems which nevertheless
contain the essential features of quantum mechanics.
We present a Haskell-based method for learning quantum mechanics that takes
place within a senior-level quantum mechanics course.
Students in the course may have no experience with Haskell
or programming at all.
We take the attitude of Papert\cite{papert}
and others\cite{sicm,sussmanFDG,alegreTFPIE2015,walck2014}
that students are aided in their learning by having
building blocks with which to create interesting structures,
that such creative activity is a motivating and effective
way to learn, and that the feedback provided by
computer-language-based building blocks can expose our
confusions and produce delight in our achievements.

The paper is organized as follows.
In section~\ref{lablang}, we introduce a laboratory language
for the description of experiments with spin-1/2 particles.
In section~\ref{calclang}, we describe a calculational language
for working with kets and operators, the abstract objects
used to do calculations.
In section~\ref{implablang}, we describe a simplified laboratory
language that students are asked to implement using the
calculational language.

\section{Laboratory Language}
\label{lablang}

The Stern-Gerlach experiment was performed
by Otto Stern and Walther Gerlach in 1922, a time
when the theory of quantum mechanics was being developed.
The experiment demonstrates the quantization of angular
momentum, which we now know comes in integer and half-integer
multiples of $\hbar$, Planck's constant.
Several modern quantum mechanics textbooks begin the subject
with the Stern-Gerlach experiment.\cite{schumacherwestmoreland,townsend,sakurainapolitano}
The Feynman Lectures also introduce the Stern-Gerlach experiment
early in the volume on quantum mechanics.\cite{feynmanlectures3}

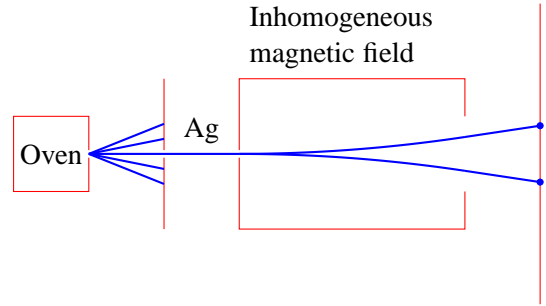
\begin{figure}
  \begin{center}
    \begin{tikzpicture}
      \draw[red] (0,0.05) -- (0,0.5) -- (-1,0.5) -- (-1,-0.5) -- (0,-0.5) -- (0,-0.05);
      \node[align=center] at (-0.5,0) {Oven};
      \draw[red] (1,0.05) -- (1,1);
      \draw[red] (1,-0.05) -- (1,-1);
      \draw[thick, blue] (0,0) -- (1,0.4);
      \draw[thick, blue] (0,0) -- (1,0.2);
      \draw[thick, blue] (0,0) -- (1,-0.2);
      \draw[thick, blue] (0,0) -- (1,-0.4);
      \draw[thick, blue] (0,0) -- (2,0);
      \node[above] at (1.5,0) {Ag};
      \draw[red] (2,0.05) -- (2,1) -- (5,1) -- (5,0.5);
      \node[align=left,above right] at (2,1) {Inhomogeneous\\magnetic field};
      \draw[red] (2,-0.05) -- (2,-1) -- (5,-1) -- (5,-0.5);
      \draw[thick, blue, domain=2:5] plot (\x, {0.025*(\x-2)*(\x-2)});
      \draw[thick, blue, domain=2:5] plot (\x, {-0.025*(\x-2)*(\x-2)});
      \draw[red] (6,-2) -- (6,2);
      \draw[thick, blue] (5,0.225) -- (6,0.375);
      \draw[thick, blue] (5,-0.225) -- (6,-0.375);
      \draw[blue,fill=blue] (6,0.375) circle [radius=0.04];
      \draw[blue,fill=blue] (6,-0.375) circle [radius=0.04];
    \end{tikzpicture}
  \end{center}
\caption{The Stern-Gerlach experiment.}
\label{sterngerlach}
\end{figure}

A schematic of the Stern-Gerlach experiment is shown in Figure~\ref{sterngerlach}.
Silver (Ag) atoms are heated in an oven and made into a beam by passing
through small holes.  Stern and Gerlach chose silver because, with a single
electron in its outer shell, it mimics the magnetic behavior of an electron.
The electron and the silver atom each possess a \emph{magnetic dipole moment},
that is, they behave like tiny bar magnets in the presence of a magnetic field.
Such a magnetic dipole moment will feel a torque, an urge to rotate, in the presence
of a magnetic field, and will moreover feel a force if the magnetic field
is \emph{inhomogeneous}, changing from one position in space to another.

The Stern-Gerlach experiment aims to produce a force on the silver atoms
with an inhomogeneous magnetic field oriented in a particular direction,
say the $z$~direction.  The silver atoms in the beam are then expected
to deflect upward or downward in the $z$~direction, depending on the extent
to which their magnetic dipole moments (vectors oriented along the imagined
tiny bar magnets) point in the negative or positive $z$~direction.
Classical (pre-quantum) physics predicts that the random distribution of
magnetic dipole moments coming from the oven should produce a continuous
spectrum of deflection of the silver atom beam.

Instead, what is seen in the experiment is that the beam splits into
two beams, and produces two spots on a detecting screen.
Gerlach called this ``directional quantization''\cite{townsend},
and, since magnetic dipole moment is proportional to angular momentum,
we now think of this as quantization of angular momentum.
The electron and the silver atom are called spin-1/2 particles
because the $z$-component of
angular momentum for particles in one of the two beams is $(1/2)\hbar$,
and that in other beam is $-(1/2)\hbar$.
Spin-1/2 particles are particles that have two outcomes in a Stern-Gerlach
experiment (as such, they are examples of quantum bits, or qubits).
Although we won't talk about them in this paper, it may be helpful
to know that spin-1 particles have three outcomes in a Stern-Gerlach type
experiment, spin-3/2 particles have four outcomes, and so on.

Our aim is to use Stern-Gerlach technology to split beams, recombine beams,
and act on single beams with uniform magnetic fields.
For this purpose, we now introduce the central data type in the laboratory
language, the \DataTypeTok{\texttt{BeamStack}}.
As in an RPN (reverse Polish notation) calculator, the data type maintains
a stack of beams to be acted on in various ways.
Figure~\ref{expapicore} shows \DataTypeTok{\texttt{BeamStack}} as
an opaque data type.  Since the primary pedagogical purpose of the laboratory
language is to use it to explore what can happen in experimental setups
that the user can design, the focus is not on the implementation
of the \DataTypeTok{\texttt{BeamStack}} data type.
Also shown in Figure~\ref{expapicore} is an initial \DataTypeTok{\texttt{BeamStack}},
called \FunctionTok{\texttt{randomBeam}},
for the single beam coming out of the oven, and a couple of utility
functions to manipulate the stack.
These functions and the rest of the laboratory language
are available in the module \texttt{Physics.Learn.BeamStack} in the \emph{learn-physics}
package\cite{learn-physics}.

\begin{figure}
\begin{Highlighting}[]
\KeywordTok{data} \DataTypeTok{BeamStack}
\FunctionTok{randomBeam} \SpecialCharTok{::} \DataTypeTok{BeamStack}
\FunctionTok{dropBeam}   \SpecialCharTok{::} \DataTypeTok{BeamStack} \SpecialCharTok{->} \DataTypeTok{BeamStack}
\FunctionTok{flipBeams}  \SpecialCharTok{::} \DataTypeTok{BeamStack} \SpecialCharTok{->} \DataTypeTok{BeamStack}
\end{Highlighting}
\caption{The \DataTypeTok{\texttt{BeamStack}} data type is a collection of beams
  organized into a stack.  The stack consisting of a single beam coming out of an oven
  is called \FunctionTok{\texttt{randomBeam}}.
  The function \FunctionTok{\texttt{dropBeam}} removes the top beam from the stack.
  The function \FunctionTok{\texttt{flipBeams}} interchanges the order of the top two beams
  on the stack.
}
\label{expapicore}
\end{figure}

We can learn quite a bit more from the Stern-Gerlach experiment
if we can do sequential Stern-Gerlach measurements, that is if we can take
one of the outcoming beams from the inhomogeneous magnetic field
and send it into another Stern-Gerlach device.  For this purpose
it is helpful to have a Stern-Gerlach splitter that creates two
parallel beams.  Such a device is shown in Figure~\ref{sgsplitter}(a).
The key to making the beams parallel is to put an oppositely
directed inhomogeneous magnetic field immediately after the
first field to deflect the beams back to parallel.

\begin{figure}
\begin{center}
    \begin{tikzpicture}
      \draw[thick, blue] (0,0) -- (2,0);
      \draw[red] (2,0.5) -- (2,1) -- (5,1) -- (5,0.5);
      \node[align=left,above right] at (2,1) {Inhomogeneous\\magnetic field};
      \draw[red] (2,-0.5) -- (2,-1) -- (5,-1) -- (5,-0.5);
      \draw[red] (5,0.5) -- (5,1) -- (8,1) -- (8,0.7);
      \node[align=left,above right] at (5,1) {Opposite\\magnetic field};
      \draw[red] (5,-0.5) -- (5,-1) -- (8,-1) -- (8,-0.7);
      \draw[thick, blue, domain=2:5] plot (\x, { 0.025*(\x-2)*(\x-2)});
      \draw[thick, blue, domain=2:5] plot (\x, {-0.025*(\x-2)*(\x-2)});
      \draw[thick, blue, domain=5:8] plot (\x, { 0.025*(18-(\x-8)*(\x-8))});
      \draw[thick, blue, domain=5:8] plot (\x, {-0.025*(18-(\x-8)*(\x-8))});
      \draw[thick, blue] (8,0.45) -- (10,0.45);
      \draw[thick, blue] (8,-0.45) -- (10,-0.45);
      \draw[->] (10.5,-0.75) -- (10.5,0.75);
      \node[right] at (10.5,0) {$z$};
    \end{tikzpicture}

(a)
\end{center}
  \begin{center}
    \begin{tikzpicture}
      \draw[thick, blue] (0,0) -- (2,0);
      \draw[red] (2,-0.75) -- (2,0.75) -- (4,0.75) -- (4,-0.75) -- cycle;
      \node[align=center] at (3,0) {\FunctionTok{\texttt{splitZ}}};
      \draw[thick, blue] (4,0.5) -- (6,0.5);
      \draw[thick, blue] (4,-0.5) -- (6,-0.5);
    \end{tikzpicture}

(b)
\end{center}
\begin{center}
\begin{Highlighting}[]
\FunctionTok{splitX} \SpecialCharTok{::} \DataTypeTok{BeamStack} \SpecialCharTok{->} \DataTypeTok{BeamStack}
\FunctionTok{splitY} \SpecialCharTok{::} \DataTypeTok{BeamStack} \SpecialCharTok{->} \DataTypeTok{BeamStack}
\FunctionTok{splitZ} \SpecialCharTok{::} \DataTypeTok{BeamStack} \SpecialCharTok{->} \DataTypeTok{BeamStack}
\FunctionTok{split}  \SpecialCharTok{::} \DataTypeTok{Radians} \SpecialCharTok{->} \DataTypeTok{Radians} \SpecialCharTok{->} \DataTypeTok{BeamStack} \SpecialCharTok{->} \DataTypeTok{BeamStack}
\end{Highlighting}

(c)
  \end{center}
\caption{The Stern-Gerlach beam splitter.
  (a) A Stern-Gerlach splitter oriented in the $z$~direction.
  (b) Schematic representation of the splitter in the $z$~direction,
   using the \FunctionTok{\texttt{splitZ}} function from part (c)
   of the figure.
  (c) Laboratory language functions for Stern-Gerlach
  beam splitters oriented in the $x$, $y$, and $z$ directions.
  The \FunctionTok{\texttt{split}} function takes two spherical coordinates
  in radians as
  arguments so that the splitting can be done in an arbitrary direction.
  (\DataTypeTok{\texttt{Radians}} is a type synonym for \DataTypeTok{\texttt{Double}}.)
  These functions act on the top (most recent) beam of the stack, remove that
  beam from the stack, and replace it with two new beams.}
\label{sgsplitter}
\end{figure}

A schematic view of the SG (Stern-Gerlach) beam splitter is shown in
Figure~\ref{sgsplitter}(b), which is labeled with the laboratory language
function \FunctionTok{\texttt{splitZ}}.  The \FunctionTok{\texttt{splitZ}}
function takes a \DataTypeTok{\texttt{BeamStack}} as input, pops the top
beam off of the stack, and replaces it with two new beams.  The lower beam
on the right side of the splitter (the \emph{spin-down} beam)
is placed on the top of the stack.  Figure~\ref{sgsplitter}(c) lists
functions for beam splitters in various directions.

The quantum mechanics book by Townsend\cite{townsend} gives a sequence
of SG experiments that help to show what a theory of quantum mechanics
needs to explain, or at least predict.  Townsend's Experiment~1
is designed to show that although there is randomness in the measurement
of spin-1/2 particles, there is not complete randomness in every measurement.
In Experiment 1, shown in Figure~\ref{townsend1fig},
the $z$-spin-up beam of the first SG splitter goes into a second SG splitter
oriented in the same direction.
The results show that when a beam of $z$-spin-up particles
enter a $z$-splitter, the entire beam comes out with $z$-spin-up.
The intensity of $0.0$ in the $z$-spin-down beam coming out of the second
splitter represents a beam of no particles, or no beam at all.
Part~(b) of Figure~\ref{townsend1fig} shows the use of the laboratory
language in GHCi to carry out Experiment 1.  The stack is shown so that
the top beam on the stack is printed last, or at the bottom of the printed
list.  We use the \FunctionTok{\texttt{dropBeam}} function
because we have no further use for the $z$-spin-down beam exiting
the first splitter.  It is not necessary to drop the beam; we could
have flipped the beams instead to act with the second splitter
on the beam we want while continuing to include all beams in the stack.

\begin{figure}
  \begin{center}
    \begin{tikzpicture}
      \draw[thick, blue] (0,0) -- (2,0);
      \node[above] at (1,0) {$1.0$};
      \draw[red] (2,-0.75) -- (2,0.75) -- (4,0.75) -- (4,-0.75) -- cycle;
      \node[align=center] at (3,0) {\FunctionTok{\texttt{splitZ}}};
      \draw[thick, blue] (4,0.5) -- (6,0.5);
      \node[above] at (5,0.5) {$0.5$};
      \draw[thick, blue] (4,-0.5) -- (5,-0.5);
      \node[above] at (4.5,-0.5) {$0.5$};
      \draw[red] (5,-0.75) -- (5,-0.25);
      \draw[red] (6,-0.25) -- (6,1.25) -- (8,1.25) -- (8,-0.25) -- cycle;
      \node[align=center] at (7,0.5) {\FunctionTok{\texttt{splitZ}}};
      \draw[thick, blue] (8,1) -- (10,1);
      \node[above] at (9,1) {$0.5$};
      \draw[thick, blue] (8,0) -- (10,0);
      \node[above] at (9,0) {$0.0$};
    \end{tikzpicture}

(a)
  \end{center}
\begin{center}
\begin{Highlighting}[]
GHCi, version 7.10.2: http://www.haskell.org/ghc/  :? for help
Prelude> \color{blue}{:m Physics.Learn.BeamStack}
Prelude Physics.Learn.BeamStack> \color{blue}{randomBeam}
\darkgreen{Beam of intensity 1.0}
Prelude Physics.Learn.BeamStack> \color{blue}{splitZ it}
\darkgreen{Beam of intensity 0.5}
\darkgreen{Beam of intensity 0.5}
Prelude Physics.Learn.BeamStack> \color{blue}{dropBeam it}
\darkgreen{Beam of intensity 0.5}
Prelude Physics.Learn.BeamStack> \color{blue}{splitZ it}
\darkgreen{Beam of intensity 0.5}
\darkgreen{Beam of intensity 0.0}
\end{Highlighting}

(b)
\end{center}
\caption{Townsend's\cite{townsend} Experiment 1.  (a) Measuring the same thing twice
  in succession gives the same results.  Every particle that is found to deflect
  in the positive $z$ direction at the first splitter also deflects in the
  positive $z$ direction at the second splitter.  We see this from the intensities.
  The entire intensity of $0.5$ that enters the second splitter comes out with
  positive deflection.  The intensity of $0.0$ in the negatively deflected beam
  means that no particles are deflected in the negative $z$ direction at the
  second splitter.
  (b) A GHCi transcript showing use of the laboratory language to obtain the
  results of Experiment 1.  The beam at the top of the stack is the
  last beam printed and hence appears at the bottom of the list.}
\label{townsend1fig}
\end{figure}

In Experiment 1, a combination of splitting and dropping occurs
that is called \emph{filtering}.  The first splitter is used
to filter the beam for particles that have spin-up in the $z$~direction.
The filtering operation happens often enough that it is useful to name
it.  Figure~\ref{expapifilter} shows several filtering functions
that we will use in upcoming experiments.

\begin{figure}
\begin{Highlighting}[]
\FunctionTok{xpFilter} \SpecialCharTok{::} \DataTypeTok{BeamStack} \SpecialCharTok{->} \DataTypeTok{BeamStack}
\FunctionTok{xpFilter} \OperatorTok{=} \FunctionTok{dropBeam} \OperatorTok{.} \FunctionTok{splitX}

\FunctionTok{xmFilter} \SpecialCharTok{::} \DataTypeTok{BeamStack} \SpecialCharTok{->} \DataTypeTok{BeamStack}
\FunctionTok{xmFilter} \OperatorTok{=} \FunctionTok{dropBeam} \OperatorTok{.} \FunctionTok{flipBeams} \OperatorTok{.} \FunctionTok{splitX}

\FunctionTok{zpFilter} \SpecialCharTok{::} \DataTypeTok{BeamStack} \SpecialCharTok{->} \DataTypeTok{BeamStack}
\FunctionTok{zpFilter} \OperatorTok{=} \FunctionTok{dropBeam} \OperatorTok{.} \FunctionTok{splitZ}

\FunctionTok{zmFilter} \SpecialCharTok{::} \DataTypeTok{BeamStack} \SpecialCharTok{->} \DataTypeTok{BeamStack}
\FunctionTok{zmFilter} \OperatorTok{=} \FunctionTok{dropBeam} \OperatorTok{.} \FunctionTok{flipBeams} \OperatorTok{.} \FunctionTok{splitZ}
\end{Highlighting}
\caption{Filtering is the composition of splitting and dropping.
  The function \FunctionTok{\texttt{xpFilter}} keeps only the beam that deflected in the positive $x$ direction.
  Since the beam that deflected in the negative $x$ direction is placed on the top of the stack
  in the \FunctionTok{\texttt{split}} function, we merely have to drop it.
  The function \FunctionTok{\texttt{xmFilter}} keeps only the beam that deflected in the negative $x$ direction.
  Since the beam that deflected in the negative $x$ direction is placed on the top of the stack
  in the \FunctionTok{\texttt{split}} function, we need to flip the beams on the stack so that we
  drop the positive $x$ beam.
  The functions \FunctionTok{\texttt{ypFilter}} and \FunctionTok{\texttt{ymFilter}} could,
  of course, also be defined.
}
\label{expapifilter}
\end{figure}

In Townsend's Experiment 2, shown in Figure~\ref{townsend2fig},
the second $z$-splitter of Experiment 1 is replaced by an $x$-splitter.
We see that an incoming beam of $z$-spin-up particles experiences
a 50/50~split at the $x$-splitter.
Although it is not shown in the figure,
an incoming beam of $z$-spin-down particles will also experience
a 50/50~split at an $x$-splitter.

\begin{figure}
  \begin{center}
    \begin{tikzpicture}
      \draw[thick, blue] (0,0) -- (2,0);
      \node[above] at (1,0) {$1.0$};
      \draw[red] (2,-0.75) -- (2,0.75) -- (4,0.75) -- (4,-0.75) -- cycle;
      \node[align=center] at (3,0) {\FunctionTok{\texttt{splitZ}}};
      \draw[thick, blue] (4,0.5) -- (6,0.5);
      \node[above] at (5,0.5) {$0.5$};
      \draw[thick, blue] (4,-0.5) -- (5,-0.5);
      \node[above] at (4.5,-0.5) {$0.5$};
      \draw[red] (5,-0.75) -- (5,-0.25);
      \draw[red] (6,-0.25) -- (6,1.25) -- (8,1.25) -- (8,-0.25) -- cycle;
      \node[align=center] at (7,0.5) {\FunctionTok{\texttt{splitX}}};
      \draw[thick, blue] (8,1) -- (10,1);
      \node[above] at (9,1) {$0.25$};
      \draw[thick, blue] (8,0) -- (10,0);
      \node[above] at (9,0) {$0.25$};
    \end{tikzpicture}

(a)
  \end{center}
  \begin{center}
    \begin{tikzpicture}
      \draw[thick, blue] (0,0) -- (2,0);
      \node[above] at (1,0) {$1.0$};
      \draw[red] (2,-0.75) -- (2,0.75) -- (4,0.75) -- (4,-0.75) -- cycle;
      \node[align=center] at (3,0) {\FunctionTok{\texttt{zpFilter}}};
      \draw[thick, blue] (4,0) -- (6,0);
      \node[above] at (5,0) {$0.5$};
      \draw[red] (6,-0.75) -- (6,0.75) -- (8,0.75) -- (8,-0.75) -- cycle;
      \node[align=center] at (7,0) {\FunctionTok{\texttt{splitX}}};
      \draw[thick, blue] (8,0.5) -- (10,0.5);
      \node[above] at (9,0.5) {$0.25$};
      \draw[thick, blue] (8,-0.5) -- (10,-0.5);
      \node[above] at (9,-0.5) {$0.25$};
    \end{tikzpicture}

(b)
  \end{center}
\begin{center}
\begin{Highlighting}[]
GHCi, version 7.10.2: http://www.haskell.org/ghc/  :? for help
Prelude> \color{blue}{:m Physics.Learn.BeamStack}
Prelude Physics.Learn.BeamStack> \color{blue}{zpFilter randomBeam}
\darkgreen{Beam of intensity 0.5}
Prelude Physics.Learn.BeamStack> \color{blue}{splitX it}
\darkgreen{Beam of intensity 0.25000000000000006}
\darkgreen{Beam of intensity 0.24999999999999994}
\end{Highlighting}

(c)
\end{center}
\caption{Townsend's Experiment 2.
  (a) Schematic diagram with splitters.
  (b) Alternate diagram of the same experiment using a filter.
  (c) GHCi transcript.
}
\label{townsend2fig}
\end{figure}

Townsend's Experiment 3 extends Experiment 2 with a third
splitter, so that the orientations of the splitters are $z$
then $x$ then $z$.  At the output of the last splitter,
we now have equal intensities of $z$-spin-up and $z$-spin-down
beams.
This result may be surprising or puzzling
when compared with Experiment 1, in which a second $z$-splitter
sends all of the particles in the direction in which they
split at a previous $z$-splitter.
In Experiment 3, all of the particles entering the last $z$-splitter
had previously split upward at the first $z$-splitter, yet now half
of those entering the last $z$-splitter are splitting downward.
Experiment 3 can also be viewed as inserting a filter for $x$-spin-up
particles between the splitters of Experiment 1.
Clearly this filter is having a significant and strange effect
on the final splitting.  It seems as though the particles have
``forgotten'' that they had previously split upwards at a $z$-splitter.
This is a crucial observation that will need to be reflected in the
theory.

Experiment 3 shows that the upper beam exiting the $x$-splitter
will undergo a 50/50 split at the subsequent $z$-splitter.
It is also the case, as can be tested with the laboratory language,
that the lower beam exiting the $x$-splitter
will also undergo a 50/50 split if sent into a $z$-splitter.

\begin{figure}
  \begin{center}
    \begin{tikzpicture}
      \draw[thick, blue] (0,0) -- (2,0);
      \node[above] at (1,0) {$1.0$};
      \draw[red] (2,-0.75) -- (2,0.75) -- (4,0.75) -- (4,-0.75) -- cycle;
      \node[align=center] at (3,0) {\FunctionTok{\texttt{splitZ}}};
      \draw[thick, blue] (4,0.5) -- (6,0.5);
      \node[above] at (5,0.5) {$0.5$};
      \draw[thick, blue] (4,-0.5) -- (5,-0.5);
      \node[above] at (4.5,-0.5) {$0.5$};
      \draw[red] (5,-0.75) -- (5,-0.25);
      \draw[red] (6,-0.25) -- (6,1.25) -- (8,1.25) -- (8,-0.25) -- cycle;
      \node[align=center] at (7,0.5) {\FunctionTok{\texttt{splitX}}};
      \draw[thick, blue] (8,1) -- (10,1);
      \node[above] at (9,1) {$0.25$};
      \draw[thick, blue] (8,0) -- (9,0);
      \node[above] at (8.5,0) {$0.25$};
      \draw[red] (9,-0.25) -- (9,0.25);
      \draw[red] (10,0.25) -- (10,1.75) -- (12,1.75) -- (12,0.25) -- cycle;
      \node[align=center] at (11,1) {\FunctionTok{\texttt{splitZ}}};
      \draw[thick, blue] (12,1.5) -- (14,1.5);
      \node[above] at (13,1.5) {$0.125$};
      \draw[thick, blue] (12,0.5) -- (14,0.5);
      \node[above] at (13,0.5) {$0.125$};
    \end{tikzpicture}

(a)
  \end{center}
\begin{center}
\begin{Highlighting}[]
Prelude Physics.Learn.BeamStack> \color{blue}{randomBeam}
\darkgreen{Beam of intensity 1.0}
Prelude Physics.Learn.BeamStack> \color{blue}{splitZ it}
\darkgreen{Beam of intensity 0.5}
\darkgreen{Beam of intensity 0.5}
Prelude Physics.Learn.BeamStack> \color{blue}{dropBeam it}
\darkgreen{Beam of intensity 0.5}
Prelude Physics.Learn.BeamStack> \color{blue}{splitX it}
\darkgreen{Beam of intensity 0.25000000000000006}
\darkgreen{Beam of intensity 0.24999999999999994}
Prelude Physics.Learn.BeamStack> \color{blue}{dropBeam it}
\darkgreen{Beam of intensity 0.25000000000000006}
Prelude Physics.Learn.BeamStack> \color{blue}{splitZ it}
\darkgreen{Beam of intensity 0.12500000000000006}
\darkgreen{Beam of intensity 0.125}
\end{Highlighting}

(b)
\end{center}
\caption{Townsend's Experiment 3.  Some particles split downward
  at the last $z$-splitter, even though all particles entering the
  last $z$-splitter have previously split upward at the first
  $z$-splitter.  This tempers the results of Experiment 1,
  which showed that no particles would split downward at the second $z$-splitter
  after they had split upward at the first $z$-splitter.
  Reconciling Experiments 1 and 3 is an important job for the theory.
  (a) Schematic diagram with splitters.
  (b) GHCi transcript.  There are many alternate ways to do this,
  including using filters.
}
\label{townsend3fig}
\end{figure}

We can recombine two beams with the same kind of inhomogeneous
magnetic fields that we used to split a beam.  Figure~\ref{sgrecombiner}
shows an SG recombiner.  Recombining is not as intuitive as it
might seem.  If the two beams that enter a recombiner did not
come from a splitter in the same direction, there is no guarantee
that they will bend the right way to merge them into a single beam.
For example, flipping two beams before recombining will generally
give different results (and often a beam intensity of zero)
from simply recombining the two beams.\footnote{I thank my
student Justin Cammarota for noticing this and bringing it
to my attention.}

\begin{figure}
  \begin{center}
    \begin{tikzpicture}
      \draw[thick, blue] (0,0.45) -- (2,0.45);
      \draw[thick, blue] (0,-0.45) -- (2,-0.45);
      \draw[red] (2,0.7) -- (2,1) -- (5,1) -- (5,0.5);
      \node[align=left,above right] at (2,1) {Inhomogeneous\\magnetic field};
      \draw[red] (2,-0.7) -- (2,-1) -- (5,-1) -- (5,-0.5);
      \draw[red] (5,0.5) -- (5,1) -- (8,1) -- (8,0.5);
      \node[align=left,above right] at (5,1) {Opposite\\magnetic field};
      \draw[red] (5,-0.5) -- (5,-1) -- (8,-1) -- (8,-0.5);
      \draw[thick, blue, domain=2:5] plot (\x, { 0.025*(18-(\x-2)*(\x-2))});
      \draw[thick, blue, domain=2:5] plot (\x, {-0.025*(18-(\x-2)*(\x-2))});
      \draw[thick, blue, domain=5:8] plot (\x, { 0.025*(\x-8)*(\x-8)});
      \draw[thick, blue, domain=5:8] plot (\x, {-0.025*(\x-8)*(\x-8)});
      \draw[thick, blue] (8,0) -- (10,0);
      \draw[->] (10.5,-0.75) -- (10.5,0.75);
      \node[right] at (10.5,0) {$z$};
    \end{tikzpicture}

(a)
  \end{center}
  \begin{center}
    \begin{tikzpicture}
      \draw[thick, blue] (7.5,1) -- (9.5,1);
      \draw[thick, blue] (7.5,0) -- (9.5,0);
      \draw[red] (9.5,-0.25) -- (9.5,1.25) -- (12,1.25) -- (12,-0.25) -- cycle;
      \node[align=center] at (10.75,0.5) {\FunctionTok{\texttt{recombineZ}}};
      \draw[thick, blue] (12,0.5) -- (14,0.5);
    \end{tikzpicture}

(b)
  \end{center}
\begin{center}
\begin{Highlighting}[]
\FunctionTok{recombineX} \SpecialCharTok{::} \DataTypeTok{BeamStack} \SpecialCharTok{->} \DataTypeTok{BeamStack}
\FunctionTok{recombineY} \SpecialCharTok{::} \DataTypeTok{BeamStack} \SpecialCharTok{->} \DataTypeTok{BeamStack}
\FunctionTok{recombineZ} \SpecialCharTok{::} \DataTypeTok{BeamStack} \SpecialCharTok{->} \DataTypeTok{BeamStack}
\FunctionTok{recombine}  \SpecialCharTok{::} \DataTypeTok{Radians} \SpecialCharTok{->} \DataTypeTok{Radians} \SpecialCharTok{->} \DataTypeTok{BeamStack} \SpecialCharTok{->} \DataTypeTok{BeamStack}
\end{Highlighting}

(c)
  \end{center}
\caption{The Stern-Gerlach beam recombiner.
  (a) A Stern-Gerlach recombiner oriented in the $z$ direction.
  (b) Schematic representation of the recombiner in the $z$~direction,
   using the \FunctionTok{\texttt{recombineZ}} function from part (c)
   of the figure.
  (c) Stern-Gerlach beam recombiners oriented in the $x$, $y$, and $z$ directions.
  The \FunctionTok{\texttt{recombine}} function takes two spherical coordinates in radians as
  arguments so that the recombining can be done in an arbitrary direction.
  These functions act on the top two beams of the stack, remove those
  beams from the stack, and replace them with a single new beam.
}
\label{sgrecombiner}
\end{figure}

Townsend's Experiment 4, shown in Figure~\ref{townsend4fig},
builds on Experiment 3 by adding an $x$-recombiner after the $x$-splitter.
From Experiment 3, we know that each of the beams with intensity 0.25
in Figure~\ref{townsend4fig} would experience a 50/50 split at a $z$-splitter.
Recombining the two beams before the $z$-splitter produces a 100/0 split!
The final $z$-splitter produces no $z$-spin-down particles, just as the
final $z$-splitter in Experiment~1 produced no $z$-spin-down particles.
Whereas the $x$-splitter in Experiment~3 disturbed the repeatability
of the $z$-splitter results in Experiment~1, the $x$-splitter-recombiner
combination in Experiment 4 does not disturb the repeatability.
This is another result that the theory will need to deal with.

\begin{figure}
  \begin{center}
    \begin{tikzpicture}
      \draw[thick, blue] (1,0) -- (2,0);
      \node[above] at (1.5,0) {$1.0$};
      \draw[red] (2,-0.75) -- (2,0.75) -- (4,0.75) -- (4,-0.75) -- cycle;
      \node[align=center] at (3,0) {\FunctionTok{\texttt{splitZ}}};
      \draw[thick, blue] (4,0.5) -- (6,0.5);
      \node[above] at (5,0.5) {$0.5$};
      \draw[thick, blue] (4,-0.5) -- (5,-0.5);
      \node[above] at (4.5,-0.5) {$0.5$};
      \draw[red] (5,-0.75) -- (5,-0.25);
      \draw[red] (6,-0.25) -- (6,1.25) -- (8,1.25) -- (8,-0.25) -- cycle;
      \node[align=center] at (7,0.5) {\FunctionTok{\texttt{splitX}}};
      \draw[thick, blue] (8,1) -- (9.5,1);
      \node[above] at (8.75,1) {$0.25$};
      \draw[thick, blue] (8,0) -- (9.5,0);
      \node[above] at (8.75,0) {$0.25$};
      \draw[red] (9.5,-0.25) -- (9.5,1.25) -- (12,1.25) -- (12,-0.25) -- cycle;
      \node[align=center] at (10.75,0.5) {\FunctionTok{\texttt{recombineX}}};
      \draw[thick, blue] (12,0.5) -- (13,0.5);
      \node[above] at (12.5,0.5) {$0.5$};
      \draw[red] (13,-0.25) -- (13,1.25) -- (15,1.25) -- (15,-0.25) -- cycle;
      \node[align=center] at (14,0.5) {\FunctionTok{\texttt{splitZ}}};
      \draw[thick, blue] (15,1) -- (16,1);
      \node[above] at (15.5,1) {$0.5$};
      \draw[thick, blue] (15,0) -- (16,0);
      \node[above] at (15.5,0) {$0.0$};
    \end{tikzpicture}

(a)
  \end{center}
\begin{center}
\begin{Highlighting}[]
Prelude Physics.Learn.BeamStack> \color{blue}{randomBeam}
\darkgreen{Beam of intensity 1.0}
Prelude Physics.Learn.BeamStack> \color{blue}{splitZ it}
\darkgreen{Beam of intensity 0.5}
\darkgreen{Beam of intensity 0.5}
Prelude Physics.Learn.BeamStack> \color{blue}{dropBeam it}
\darkgreen{Beam of intensity 0.5}
Prelude Physics.Learn.BeamStack> \color{blue}{splitX it}
\darkgreen{Beam of intensity 0.25000000000000006}
\darkgreen{Beam of intensity 0.24999999999999994}
Prelude Physics.Learn.BeamStack> \color{blue}{recombineX it}
\darkgreen{Beam of intensity 0.5}
Prelude Physics.Learn.BeamStack> \color{blue}{splitZ it}
\darkgreen{Beam of intensity 0.5}
\darkgreen{Beam of intensity 0.0}
\end{Highlighting}

(b)
\end{center}
\caption{Townsend's Experiment 4.  One way to view this experiment
is that an $x$-splitter-recombiner pair has been inserted into
the apparatus of Experiment 1, and the same results are obtained
as in Experiment 1.  A second way to view this experiment is that
two beams, each of which would produce a 50/50 split in a
\FunctionTok{\texttt{splitZ}}, are being recombined into a beam
that produces a 100/0 split in a \FunctionTok{\texttt{splitZ}}.
}
\label{townsend4fig}
\end{figure}

Of course we can use the laboratory language to go beyond Townsend's
experiments.  Students can invent other configurations of splitters
and recombiners and see if the results match their expectations.

The last basic building block of the laboratory language
is application of a uniform magnetic field to a beam.
A magnetic field can be applied in a particular direction
with a certain strength for a certain amount of time.
The functions to do this are as follows.
\begin{Highlighting}[]
\FunctionTok{applyBFieldX} \SpecialCharTok{::} \DataTypeTok{Radians} \SpecialCharTok{->} \DataTypeTok{BeamStack} \SpecialCharTok{->} \DataTypeTok{BeamStack}
\FunctionTok{applyBFieldY} \SpecialCharTok{::} \DataTypeTok{Radians} \SpecialCharTok{->} \DataTypeTok{BeamStack} \SpecialCharTok{->} \DataTypeTok{BeamStack}
\FunctionTok{applyBFieldZ} \SpecialCharTok{::} \DataTypeTok{Radians} \SpecialCharTok{->} \DataTypeTok{BeamStack} \SpecialCharTok{->} \DataTypeTok{BeamStack}
\FunctionTok{applyBField}  \SpecialCharTok{::} \DataTypeTok{Radians} \SpecialCharTok{->} \DataTypeTok{Radians} \SpecialCharTok{->} \DataTypeTok{Radians} \SpecialCharTok{->} \DataTypeTok{BeamStack} \SpecialCharTok{->} \DataTypeTok{BeamStack}
\end{Highlighting}
These functions apply a uniform magnetic field to the top beam of the stack.
In the function \\
\FunctionTok{\texttt{applyBFieldX}}, the \DataTypeTok{\texttt{Radians}}
argument is an angle in radians that represents a combination of the strength
of the applied magnetic field and the duration over which it is applied.
The \FunctionTok{\texttt{applyBField}} function takes two spherical coordinates as
arguments to represent the direction of the applied magnetic field, and a third
numerical argument to represent the combination of magnetic field strength and
time over which the field is applied.

Here are some example puzzles for students to work on as they
explore the laboratory language.
\begin{itemize}
\item
First, find a sequence of two filters such that no
particles exit the second filter (no particles is the
same as a beam of zero intensity).
Now, is it possible to find a third filter to place
between the first two, such that particles now
flow from the last filter?  If this is possible,
we may need to adjust our intuition about what
a filter is.
\item
Can you find a direction and duration for a uniform magnetic
field to act on a beam exiting a \FunctionTok{\texttt{zpFilter}}
so that the entire beam intensity will make it through an
\FunctionTok{\texttt{xpFilter}}?
Does this suggest a way to think about what a uniform magnetic
field does?
\item
In Townsend's Experiment 4, suppose we apply a uniform magnetic field
in the $x$~direction
to the lower beam between the $x$-splitter and $x$-recombiner.
If the duration of application of the magnetic field is zero,
the results will match that of Experiment 4.  What is the next shortest
duration when the results match again?
Is the answer surprising?
\end{itemize}

Students seemed happy to play with the laboratory language,
and some came up with interesting situations.
I gave almost no instruction on how to use GHCi or Haskell,
yet students seemed to make good progress.
That function application takes precedence over
operations such as division was not intuitive for my students.
Most surprising to me was that while students asked questions
about a variety of issues, no one sought clarification about
why they were getting errors from the compiler.
They just tried different things until something worked.
I learned that I need to be proactive about explaining
error messages.

\section{Calculational Language}
\label{calclang}

Quantum Theory claims that the state of affairs for
a particle is described by a vector in a complex vector space.
In the case of a spin-1/2
particle, the vector can describe the state of the particle as it exits
a Stern-Gerlach splitter, for example.  Paul Dirac created a notation
based on the inner product, or bracket $\langle \phi | \psi \rangle$,
calling $\bra{\phi}$ a \emph{bra} and $\ket{\psi}$ a \emph{ket}.  The bra vector
is dual to the ket vector.  Jerzy Karczmarczuk gave an early
implementation of kets in Haskell.\cite{karczmarczuk2003}
Figure~\ref{calclangkets} shows ket
vectors for spin-1/2 particles.
The calculational language
is available in the module \texttt{Physics.Learn.Ket} in the \emph{learn-physics}
package\cite{learn-physics}.

\begin{figure}
\begin{tabular}{ll}
\VERB+\KeywordTok{data} \DataTypeTok{Ket}+ \\
\VERB+\FunctionTok{xp} \SpecialCharTok{::} \DataTypeTok{Ket}+ & $\ket{x_+} = \frac{1}{\sqrt{2}} \ket{z_+} + \frac{1}{\sqrt{2}} \ket{z_-}$ \\
\VERB+\FunctionTok{xm} \SpecialCharTok{::} \DataTypeTok{Ket}+ & $\ket{x_-} = \frac{1}{\sqrt{2}} \ket{z_+} - \frac{1}{\sqrt{2}} \ket{z_-}$ \\
\VERB+\FunctionTok{yp} \SpecialCharTok{::} \DataTypeTok{Ket}+ & $\ket{y_+} = \frac{1}{\sqrt{2}} \ket{z_+} + \frac{i}{\sqrt{2}} \ket{z_-}$ \\
\VERB+\FunctionTok{ym} \SpecialCharTok{::} \DataTypeTok{Ket}+ & $\ket{y_-} = \frac{1}{\sqrt{2}} \ket{z_+} - \frac{i}{\sqrt{2}} \ket{z_-}$ \\
\VERB+\FunctionTok{zp} \SpecialCharTok{::} \DataTypeTok{Ket}+ & $\ket{z_+}$ \\
\VERB+\FunctionTok{zm} \SpecialCharTok{::} \DataTypeTok{Ket}+ & $\ket{z_-}$ \\
\VERB+\FunctionTok{np} \SpecialCharTok{::} \DataTypeTok{Radians} \SpecialCharTok{->} \DataTypeTok{Radians} \SpecialCharTok{->} \DataTypeTok{Ket}+ & $\ket{n_+(\theta,\phi)} = \cos \frac{\theta}{2} \ket{z_+} + e^{i \phi} \sin \frac{\theta}{2} \ket{z_-}$ \\
\VERB+\FunctionTok{nm} \SpecialCharTok{::} \DataTypeTok{Radians} \SpecialCharTok{->} \DataTypeTok{Radians} \SpecialCharTok{->} \DataTypeTok{Ket}+ & $\ket{n_-(\theta,\phi)} = \sin \frac{\theta}{2} \ket{z_+} - e^{i \phi} \cos \frac{\theta}{2} \ket{z_-}$
\end{tabular}
\caption{Kets for spin-1/2 particles.}
\label{calclangkets}
\end{figure}

Physical quantities
such as position, momentum, angular momentum, and energy
are represented by linear operators in quantum mechanics.
For a spin-1/2 particle, the observables of interest are
components of angular momentum in various directions.
We denote by $S_x$, $S_y$, and $S_z$ the $x$-, $y$-, and
$z$-components of spin angular momentum.
The Pauli operators $\sigma_x$, $\sigma_y$, and $\sigma_z$
are then defined by $S_x = \frac{\hbar}{2} \sigma_x$,
$S_y = \frac{\hbar}{2} \sigma_y$, and
$S_z = \frac{\hbar}{2} \sigma_z$.
These Pauli operators are implemented as
\FunctionTok{\texttt{sx}},
\FunctionTok{\texttt{sy}}, and
\FunctionTok{\texttt{sz}} in Figure~\ref{calclangops}.

The power of Dirac's notation is in the Dirac product,
which is associative.
As shown in the equation below, the expression
$\ket{x_+} \langle x_+ | y_+ \rangle$
can be regarded either as an operator acting on
a ket (as shown at the left end of the equation,
where the Dirac product of a ket and a bra is an \emph{outer product},
producing an operator), or as the scalar product of a ket
and an inner product (as shown at the right end of the equation).
\[
\left( \ket{x_+} \bra{x_+} \right) \ket{y_+} = \ket{x_+} \langle x_+ | y_+ \rangle = \ket{x_+} (\langle x_+ | y_+ \rangle)
\]
We use the Haskell notation \OperatorTok{\texttt{<>}} for the Dirac product.
Figure~\ref{products} shows types for which the Dirac product is defined,
and gives examples of its use.
There is also an adjoint operation, represented by a dagger,
which turns kets into bras, bras into kets, complex numbers into
their complex conjugates, and operators into their adjoint operators.

\begin{figure}
\begin{tabular}{ll}
\VERB+\KeywordTok{data} \DataTypeTok{Operator}+ \\
\VERB+\FunctionTok{sx} \SpecialCharTok{::} \DataTypeTok{Operator}+ & $\ket{x_+} \bra{x_+} - \ket{x_-} \bra{x_-}$ \\
\VERB+\FunctionTok{sy} \SpecialCharTok{::} \DataTypeTok{Operator}+ & $\ket{y_+} \bra{y_+} - \ket{y_-} \bra{y_-}$ \\
\VERB+\FunctionTok{sz} \SpecialCharTok{::} \DataTypeTok{Operator}+ & $\ket{z_+} \bra{z_+} - \ket{z_-} \bra{z_-}$ \\
\VERB+\FunctionTok{sn} \SpecialCharTok{::} \DataTypeTok{Radians} \SpecialCharTok{->} \DataTypeTok{Radians} \SpecialCharTok{->} \DataTypeTok{Operator}+ & $\ket{n_+(\theta,\phi)} \bra{n_+(\theta,\phi)} - \ket{n_-(\theta,\phi)} \bra{n_-(\theta,\phi)}$
\end{tabular}
\caption{Operators for spin-1/2 particles.}
\label{calclangops}
\end{figure}

\begin{figure}
\begin{center}
  \begin{tabular}{c|ccc}
    Dirac       & $\ket{y_+}$ & $\bra{y_+}$ & $B$ \\
    product     & \VERB+\DataTypeTok{Ket}+ & \VERB+\DataTypeTok{Bra}+ & \VERB+\DataTypeTok{Operator}+ \\ \hline
                & \\
    $\ket{x_+}$ & $\ket{x_+} \ket{y_+}$ & $\ket{x_+} \bra{y_+}$ & $\ket{x_+} B$ \\
  \VERB+\DataTypeTok{Ket}+             & nonsense             & \VERB+\DataTypeTok{Operator}+ & nonsense \\
                & \\
    $\bra{x_+}$ & $\langle x_+ | y_+ \rangle$ & $\bra{x_+} \bra{y_+}$ & $\bra{x_+} B$  \\
  \VERB+\DataTypeTok{Bra}+ & \VERB+\DataTypeTok{C}+ & nonsense & \VERB+\DataTypeTok{Bra}+ \\
                & \\
    $A$         & $A \ket{y_+}$ & $A \bra{y_+}$ & $A B$ \\
  \VERB+\DataTypeTok{Operator}+ & \VERB+\DataTypeTok{Ket}+ & nonsense & \VERB+\DataTypeTok{Operator}+
  \end{tabular}

(a)
\end{center}
\begin{center}
\begin{tabular}{lll}
Term & Type & Notation \\ \hline
\VERB+\FunctionTok{dagger} \FunctionTok{ym} \OperatorTok{<>} \FunctionTok{xp}+
  & \VERB+\DataTypeTok{C}+
  & $\langle y_- | x_+ \rangle$ \\
\VERB+\FunctionTok{yp} \OperatorTok{<>} \FunctionTok{dagger} \FunctionTok{ym}+
  & \VERB+\DataTypeTok{Operator}+
  & $\ket{y_+} \bra{y_-}$ \\
\VERB+\FunctionTok{sx} \OperatorTok{<>} \FunctionTok{yp}+
  & \VERB+\DataTypeTok{Ket}+
  & $\sigma_x \ket{y_+}$ \\
\VERB+\FunctionTok{xp} \OperatorTok{<>} \FunctionTok{dagger} \FunctionTok{xp} \OperatorTok{<>} \FunctionTok{yp}+
  & \VERB+\DataTypeTok{Ket}+
  & $\ket{x_+} \langle x_+ | y_+ \rangle$ \\
\VERB+\FunctionTok{dagger} \FunctionTok{zm} \OperatorTok{<>} \FunctionTok{sy} \OperatorTok{<>} \FunctionTok{yp}+
  & \VERB+\DataTypeTok{C}+
  & $\bra{z_-} \sigma_y \ket{y_+}$
\end{tabular}

(b)
\end{center}
\caption{The Dirac product.  (a)  Table showing which products make sense and which do not.
  The type \DataTypeTok{\texttt{C}} is a complex number.
  (b)  Examples of the Dirac product.
  The first line is an inner product.
  The second line is an outer product.
  The function \FunctionTok{\texttt{dagger}} is an adjoint operation
  that turns kets into bras, bras into kets, operators into
  (adjoint) operators, and complex numbers into their complex conjugates.
  The Dirac product \OperatorTok{\texttt{<>}} is used for scalar
  products, inner products, outer products, operator products, and wherever
  it makes sense.
}
\label{products}
\end{figure}

The Dirac product is implemented with multi-parameter type classes
and functional dependencies.  As shown in Figure~\ref{impdirac},
the Dirac product \OperatorTok{\texttt{<>}} is owned by the type class
\DataTypeTok{\texttt{Mult}}.  The Dirac product is used to multiply
complex numbers, kets, bras, and operators.  Of the 16 pairs of
these four types, 12 make sense to multiply; each of these 12 corresponds
to an instance of type class \DataTypeTok{\texttt{Mult}}, three of
which are shown in Figure~\ref{impdirac}.  The four pairs that
make no sense to multiply are shown in Figure~\ref{products}(a).

\begin{figure}
\begin{Highlighting}[]
\KeywordTok{class} \DataTypeTok{Mult} \FunctionTok{a} \FunctionTok{b} \FunctionTok{c} \OperatorTok{|} \FunctionTok{a} \FunctionTok{b} \SpecialCharTok{->} \FunctionTok{c} \KeywordTok{where}
    \NormalTok{(}\OperatorTok{<>}\NormalTok{) }\SpecialCharTok{::} \FunctionTok{a} \SpecialCharTok{->} \FunctionTok{b} \SpecialCharTok{->} \FunctionTok{c}

\KeywordTok{instance} \DataTypeTok{Mult} \DataTypeTok{Bra} \DataTypeTok{Ket} \DataTypeTok{C} \KeywordTok{where}
    \DataTypeTok{Bra} \FunctionTok{matrixBra} \OperatorTok{<>} \DataTypeTok{Ket} \FunctionTok{matrixKet}
        \OperatorTok{=} \FunctionTok{sum} \OperatorTok{$} \FunctionTok{zipWith} \NormalTok{(}\OperatorTok{*}\NormalTok{) (}\FunctionTok{toList} \FunctionTok{matrixBra}\NormalTok{)}
                            \NormalTok{(}\FunctionTok{toList} \FunctionTok{matrixKet}\NormalTok{)}

\KeywordTok{instance} \DataTypeTok{Mult} \DataTypeTok{Operator} \DataTypeTok{Ket} \DataTypeTok{Ket} \KeywordTok{where}
    \DataTypeTok{Operator} \FunctionTok{matrixOp} \OperatorTok{<>} \DataTypeTok{Ket} \FunctionTok{matrixKet}
        \OperatorTok{=} \DataTypeTok{Ket} \NormalTok{(}\FunctionTok{matrixOp} \OperatorTok{#>} \FunctionTok{matrixKet}\NormalTok{)}

\KeywordTok{instance} \DataTypeTok{Mult} \DataTypeTok{Operator} \DataTypeTok{Operator} \DataTypeTok{Operator} \KeywordTok{where}
    \DataTypeTok{Operator} \FunctionTok{m1} \OperatorTok{<>} \DataTypeTok{Operator} \FunctionTok{m2} \OperatorTok{=} \DataTypeTok{Operator} \NormalTok{(}\FunctionTok{m1} \OperatorTok{M.<>} \FunctionTok{m2}\NormalTok{)}
\end{Highlighting}
\caption{Implementation of the Dirac product.  The multi-parameter type class \DataTypeTok{\texttt{Mult}}
  owns the Dirac product \OperatorTok{\texttt{<>}}.  Three of the twelve instances of \DataTypeTok{\texttt{Mult}}
  are shown.  The first instance is for forming the inner product of a bra and a ket,
  resulting in a complex number.
  The second instance is for an operator acting on a ket, producing another ket.
  The third instance is the operator product.
  }
\label{impdirac}
\end{figure}

In the absence of intervention, a state vector $\ket{\psi(t)}$ evolves according to the
Schr\"odinger equation,
\[
i \hbar \frac{d}{dt} \ket{\psi(t)} = H \ket{\psi(t)} ,
\]
where $H$ is a special operator called the \emph{Hamiltonian},
which describes the particle
and its interaction with the environment.
The calculational language provides a function
\begin{Highlighting}[]
\FunctionTok{timeEv} \SpecialCharTok{::} \DataTypeTok{TimeStep} \SpecialCharTok{->} \DataTypeTok{Operator} \SpecialCharTok{->} \DataTypeTok{Ket} \SpecialCharTok{->} \DataTypeTok{Ket}
\end{Highlighting}
to numerically solve the Schr\"odinger equation, using an algorithm
from \emph{Numerical Recipes}\cite{numericalrecipes}.
This function takes a time step, a Hamiltonian operator, and the current
state ket, and returns the state ket advanced by the time step.

One of the most important calculations in quantum mechanics
is finding the probability that a particular measurement
result will occur.
There are many situations in which a measurement result is associated
with an \emph{outcome ket} $\ket{\phi}$.
In these situations, the probability of the
outcome is given by the square of the magnitude of the inner product
of the outcome ket with the state ket $\ket{\psi}$.
\[
P = \abs{\langle \phi | \psi \rangle}^2
\]
In the calculational language, we would express this as follows.
\begin{Highlighting}[]
\FunctionTok{magnitude} \NormalTok{(}\FunctionTok{dagger} \FunctionTok{phi} \OperatorTok{<>} \FunctionTok{psi}\NormalTok{) }\OperatorTok{**} \DecValTok{2}
\end{Highlighting}
Here, \FunctionTok{\texttt{phi}} and \FunctionTok{\texttt{psi}} are kets,
\FunctionTok{\texttt{dagger}} \FunctionTok{\texttt{phi}} is a bra vector,
and \OperatorTok{\texttt{<>}} is the Dirac product used for scalar multiplication,
inner product (the case here), and outer product.

\section{Simplified Laboratory Language}
\label{implablang}

Having used the calculational language to solve problems,
make predictions, and do animations, we would now like to use
it to implement the laboratory language that we started with.
However, the laboratory language that we started with
requires one important feature that we have not included
in the calculational language, a feature that I do not typically
cover in a one-semester course on quantum mechanics.
What is missing is the idea of a density matrix, which is
a more general way of describing the state of a physical
system.  The only purpose to which the density matrix is put
in the laboratory language is the description of the
\FunctionTok{\texttt{randomBeam}} coming out of the oven.

Our simplified laboratory language will constrain itself
to beams that have already come out of some SG apparatus.
There is a great simplification in this constraint,
in that our central data type can now be \DataTypeTok{\texttt{Beam}},
rather than \DataTypeTok{\texttt{BeamStack}}.
The functions in Figure~\ref{simplelablang} are those that students
are asked to implement.  In terms of the \DataTypeTok{\texttt{Beam}}
data type, their type signatures are clearer and more meaningful
than the corresponding functions for working with a
\DataTypeTok{\texttt{BeamStack}}.

\begin{figure}
\begin{Highlighting}[]
\KeywordTok{data} \DataTypeTok{Beam}
\FunctionTok{xpBeam} \SpecialCharTok{::} \DataTypeTok{Beam}
\FunctionTok{xmBeam} \SpecialCharTok{::} \DataTypeTok{Beam}
\FunctionTok{ypBeam} \SpecialCharTok{::} \DataTypeTok{Beam}
\FunctionTok{ymBeam} \SpecialCharTok{::} \DataTypeTok{Beam}
\FunctionTok{zpBeam} \SpecialCharTok{::} \DataTypeTok{Beam}
\FunctionTok{zmBeam} \SpecialCharTok{::} \DataTypeTok{Beam}
\FunctionTok{intensity} \SpecialCharTok{::} \DataTypeTok{Beam} \SpecialCharTok{->} \DataTypeTok{R}
\FunctionTok{split} \SpecialCharTok{::} \DataTypeTok{Radians} \SpecialCharTok{->} \DataTypeTok{Radians} \SpecialCharTok{->} \DataTypeTok{Beam} \SpecialCharTok{->} \NormalTok{(}\DataTypeTok{Beam}\NormalTok{,}\DataTypeTok{Beam}\NormalTok{)}
\FunctionTok{splitX} \SpecialCharTok{::} \DataTypeTok{Beam} \SpecialCharTok{->} \NormalTok{(}\DataTypeTok{Beam}\NormalTok{,}\DataTypeTok{Beam}\NormalTok{)}
\FunctionTok{splitY} \SpecialCharTok{::} \DataTypeTok{Beam} \SpecialCharTok{->} \NormalTok{(}\DataTypeTok{Beam}\NormalTok{,}\DataTypeTok{Beam}\NormalTok{)}
\FunctionTok{splitZ} \SpecialCharTok{::} \DataTypeTok{Beam} \SpecialCharTok{->} \NormalTok{(}\DataTypeTok{Beam}\NormalTok{,}\DataTypeTok{Beam}\NormalTok{)}
\FunctionTok{xpFilter} \SpecialCharTok{::} \DataTypeTok{Beam} \SpecialCharTok{->} \DataTypeTok{Beam}
\FunctionTok{xmFilter} \SpecialCharTok{::} \DataTypeTok{Beam} \SpecialCharTok{->} \DataTypeTok{Beam}
\FunctionTok{zpFilter} \SpecialCharTok{::} \DataTypeTok{Beam} \SpecialCharTok{->} \DataTypeTok{Beam}
\FunctionTok{zmFilter} \SpecialCharTok{::} \DataTypeTok{Beam} \SpecialCharTok{->} \DataTypeTok{Beam}
\FunctionTok{recombine} \SpecialCharTok{::} \DataTypeTok{Radians} \SpecialCharTok{->} \DataTypeTok{Radians} \SpecialCharTok{->} \NormalTok{(}\DataTypeTok{Beam}\NormalTok{,}\DataTypeTok{Beam}\NormalTok{) }\SpecialCharTok{->} \DataTypeTok{Beam}
\FunctionTok{recombineX} \SpecialCharTok{::} \NormalTok{(}\DataTypeTok{Beam}\NormalTok{,}\DataTypeTok{Beam}\NormalTok{) }\SpecialCharTok{->} \DataTypeTok{Beam}
\FunctionTok{recombineY} \SpecialCharTok{::} \NormalTok{(}\DataTypeTok{Beam}\NormalTok{,}\DataTypeTok{Beam}\NormalTok{) }\SpecialCharTok{->} \DataTypeTok{Beam}
\FunctionTok{recombineZ} \SpecialCharTok{::} \NormalTok{(}\DataTypeTok{Beam}\NormalTok{,}\DataTypeTok{Beam}\NormalTok{) }\SpecialCharTok{->} \DataTypeTok{Beam}
\FunctionTok{applyBField} \SpecialCharTok{::} \DataTypeTok{Radians} \SpecialCharTok{->} \DataTypeTok{Radians} \SpecialCharTok{->} \DataTypeTok{Radians} \SpecialCharTok{->} \DataTypeTok{Beam} \SpecialCharTok{->} \DataTypeTok{Beam}
\FunctionTok{applyBFieldX} \SpecialCharTok{::} \DataTypeTok{Radians} \SpecialCharTok{->} \DataTypeTok{Beam} \SpecialCharTok{->} \DataTypeTok{Beam}
\FunctionTok{applyBFieldY} \SpecialCharTok{::} \DataTypeTok{Radians} \SpecialCharTok{->} \DataTypeTok{Beam} \SpecialCharTok{->} \DataTypeTok{Beam}
\FunctionTok{applyBFieldZ} \SpecialCharTok{::} \DataTypeTok{Radians} \SpecialCharTok{->} \DataTypeTok{Beam} \SpecialCharTok{->} \DataTypeTok{Beam}
\end{Highlighting}
\caption{Simplified version of the laboratory language.
  The goal is for a student to implement
  the listed functions in terms of the calculational language.
  The types \DataTypeTok{\texttt{R}} and \DataTypeTok{\texttt{Radians}}
  are synonyms for \DataTypeTok{\texttt{Double}}.
}
\label{simplelablang}
\end{figure}

Up to this point, students have not really been programming.
They have been using GHCi as a fancy calculator.
Very little needed to be explained in term of language
syntax and semantics.
At this point, some time needs to be spent on basic language
issues in order for students to be able to write function
definitions for the functions in Figure~\ref{simplelablang}.

I do not expect students to write code for the \DataTypeTok{\texttt{Beam}}
data type.  Instead, we discuss as a class the information that a
\DataTypeTok{\texttt{Beam}} needs to contain, and the options
for representing that data.  There are different ways to define
\DataTypeTok{\texttt{Beam}}; we decide as a class how we want to do it,
then I write code for \DataTypeTok{\texttt{Beam}} that the class uses
to construct the functions in Figure~\ref{simplelablang}.

Functions \FunctionTok{\texttt{xpBeam}}, \FunctionTok{\texttt{zmBeam}},
and the like are straightforward to implement once a definition
for \DataTypeTok{\texttt{Beam}} is in place.  Students had little trouble
with these functions.  The \FunctionTok{\texttt{intensity}} function
is a good next step for students in that it requires a bit of thought
and a bit of wrestling with computational details such as the difference
between a complex number (type \DataTypeTok{\texttt{C}}) that happens
to be real and a real number (type \DataTypeTok{\texttt{R}}).
The splitting functions are the real heart of the exercise.
They have been a challenge for students, but worth the effort.
I think the process of writing the splitters has helped to clarify
what the splitters are really doing.
The filter and magnetic field functions are not as hard to write
as the splitters.  The recombiners are the most difficult functions
to write.

My sense is that students experience some frustration
when they know what they want to say, but don't know how to say it
in Haskell, but in most cases they find that an idea's expression
in Haskell, once learned, seems quite reasonable.  Students appear
to experience
satisfaction when they succeed in expressing ideas like the functions in
Figure~\ref{simplelablang}, and can then use those functions
to express bigger ideas.

\end{document}